# Comformal geometric method study open strings and closed strings relationships on Dirichlet branes

Hanze Li, Maolin Bo*

Key laboratory of Extraordinary Bond Engineering and Advanced Materials Technology (EBEAM) of Chongqing,Yangtze Normal University,Chongqing 408100, China

*E-mail:bmlwd@qq.com

**Abstract**

The compactness of the closed string in the classical Type II string theory reveals the duality, whereas the compactness of the open string reveals that the end of the string is on the hypersurface which satisfies the Dirichlet boundary condition. We study the Dirichlet branes, or 'D-branes', with the concept of conformal geometry. It is further used to show that the interaction between open strings and closed strings. Also, we use the Smale's work determines that D-branes are intrinsic to Type II string theory. This method provides a deeper approach to the visualization of Type II string theory.

## 1. Introduction

The particle is a small localized object to which can be ascribed several physical or chemical properties such as brane and string. In the string theory, particle is zero-dimensional object, string is one-dimensional object and the brane is two or more higher dimensional object. There are have two important characteristics of the branes solution. First, the branes solution are non-singularity and the branes are the legally objects to exist. Second, their energy density is inversely proportional of the coupling constant and these have the weak coupling of objects. Therefore, the branes are possibility to appear in non-compacted spacetime or a space of compactified.

The existence of D-branes can be deduced from the duality of the perturbation string theory[1]. The predecessor of the string theorists has studied the properties of a series of D-branes dynamics from the duality of string theory[2]. The deeper study of D-branes has a far-reaching impact on M-theory[3]. Our understanding D-branes is using conformal geometry. The conformal geometry is a theory on the invariants of conformal transformation[4]. It's the powerful tool for studying the transformation and mapping of curved surfaces. At the same time, Smale's work has proved that the two paths in the base manifolds are regular homotopic, when and only when their ascension is on the unit tangent vector bundle[5]. The work is help to understand the open strings and closed strings in a geometric way.

In this study, we investigated the existence of D-brane, and D-brane is equivalent surface by the conformal geometry. And we also studied the interaction between open strings and closed strings. We consider the comformal geometric algebra language to calculate D-branes with Dirichlet boundary condition. And the our study will promote the development of M-theory.

## 2. Principles of differential topology of conformal geometry

In conformal geometry[4], a regular assignment of a tangent vector to each point of a manifold is called a vector field *TM*. The tangent bundle of a differentiable manifold *M* is a manifold *TM*, which assembles all the tangent vectors in *M*. For given a regular surface $S$ that all the unit tangent vectors on the surface form a three dimensional manifold unit tangent manifold $UTM(S)$[6]. Defining as:

$$UTM(S) \equiv \{(p,v) \mid p \in S, v \in TM_p(S), |v| = 1\} \tag{1}$$

Fix a point $p \in S$, all the unit tangent vectors form a circle[7]. Thus, elements of $UTM$ are pairs ($p$, $v$), where $p$ is some point of the manifold and $v$ is some tangent direction (of unit length) to the manifold at $p$. The unit tangent bundle carries a variety of differential geometric structures. Therefore, the unit tangent bundle of the surface is a fiber bundle. Where a fiber bundle is a space that is locally a product space, but globally may have a different topological structure. We construct the unit tangent bundle of the spherical surface[8].

Topologically, we use a spherical projection to set up local coordinates. We given the spherical surface of the unit $S^2 = S^1 \times S^1$. We imagine that there are a light source in the north pole and a plane cross the equator. The rays from the north pole penetrate the sphere, projecting onto the plane[9].

Which shown as the **Fig. 1**

So that the local coordinates of the spherical surface are obtained by spherical projection. Consider a point from the spherical surface ($x$, $y$, $z$) mapping onto a point from the plane ($X$, $Y$). By the calculating we obtain:

$$(X,Y) = (\frac{x}{1-z}, \frac{y}{1-z}). \tag{2}$$

However, the local coordinate is not able to represent the north pole point $p(\infty)$. So we move the light source to the south pole point, and obtain the other local coordinate ($W$, $V$):

$$(W,V) = (\frac{x}{1+z}, \frac{y}{1+z}). \tag{3}$$

consider the transform the form of $z = X + iY$, $w = W - iV$,then we have:

$$w = z^{-1}. \tag{4}$$

Also, we got the transformation formula of the cotangent vectors:

$$dw = -\frac{1}{z^2}dz. \tag{5}$$

We look at a point on the equator

$$z = e^{i\theta} \tag{6}$$

and a unit tangent vector,

$$dz = e^{i\tau}. \tag{7}$$

We take **Eq.** (6) and **Eq.** (7) transform into

$$w = e^{-i\theta} \tag{8}$$

$$dw = e^{\pi - i2\theta + \tau} \tag{9}$$

And its parameter is ($\theta$, $\tau$). We call that it is the coordinate transformation $\psi : (z, dz) \to (w, dw)$.

Now we cut the surface of spherical. The unit tangent bundle of hemispherical surface is the trivial bundle which can be represent as $\mathbb{D}^2 \times \mathbb{S}^1$ ( $\mathbb{D}^2$ is the hemispherical surface & $\mathbb{S}^1$ is the fiber). That the unit tangent bundle of hemispherical surface is a solid ring $T^2$. The unit tangent bundle of the spherical surface is equivalent to the bonding of two solid rings along the surface of the boundary[10]. The bonding mapping is the mapping

$$\psi : (\theta, \tau) \mapsto (-\theta, \pi - 2\theta + \tau) \tag{10}$$

The unit tangent bundle of the upper hemispherical surface is a solid rings, corresponding to the homotopic group

$$\pi_1(M_1) = \langle a_1 \rangle \tag{11}$$

As the same way, the solid rings on the under hemispherical surface is homotopic group

$$\pi_1(M_2) = \langle a_2 \rangle. \tag{12}$$

In terms of the Seifert-van Kampen theorem[11], we got the homotopic group of spherical unit tangent bundle as

$$\pi_1(M_1 \cup M_2) = \langle a | a^2 \rangle = \mathbb{Z}_2 \ . \tag{13}$$

Where a is fiber and $\mathbb{Z}_2$ is *Mod* 2 domain. The solution just have only two elements 0 and 1, which means that on the unit tangent of the spherical surface, all closed curves have only two homotopic classes. Futhermore, we got the more applicable inverse conclusion. If there are tow circles which are regular and homotopic so that must be a regular surface *S* to connect them [12]. Then ,we formulate

$$\partial S = \gamma_0 - \gamma_1 \tag{14}$$

Where $\gamma_0$ and $\gamma_1$ are the circles, and *S* is the regular surface[13]. In what follows though we will work towords a conformal geometic description of D-branes.

## 3. Calculation methods

### 3.1 Conformal geometry study of D-branes

We begin with specific description. By "D*p* - brane," we will mean that the *p* + 1 dimensional hyperspace surface can connect to an open string. Such an object is generated when we select the Dirichlet instead of Neumann boundary conditions[14]. More precisely, D*p*-brane is specified by selecting Neumann boundary conditions in the hypersurface direction,

$$\partial_\sigma X^i \big|_{\sigma=0,\pi} = 0, \ i = 0,1,\ldots, p. \tag{15}$$

And Dirichlet boundary conditions[15] in the transverse directions,

$$\delta x^i \big|_{\sigma=0,\pi} = 0 \ \ i = p+1,\ldots,9 . \tag{16}$$

In the type IIA string theory[16], these branes exist for all even values of *p*,

$$p = 0, 2, 4, 6, 8 .$$

The case *p*=0 is a “D-particle”, while *p*=8 describes a “domain wall” in ten dimensional[17]. The $D_0$-brane and $D_6$-brane are electromagnetic dual of each other, as are the $D_2$-brane and $D_4$-brane. In the type IIB string theory[18], branes are all odd values of *p*,

$$p = -1, 1, 3, 5, 7, 9 .$$

The case $p$ = -1 describes an object while is localized in time and corresponds to a "D-instanton", while $p$ = 1 is a "D-string", The $D_9$-branes are spacetim filling branes while p=3 yields the self-dual $D_3$-brane. The D-instanon and $D_7$-brane are electromagnetic dual of one another, as are the $D_1$-brane and the $D_5$-brane[14].

To see the nature of the dual theory[19], consider the open string boundary condition[20]

$$
\begin{aligned}
0=\partial_n = X^{\mu} &= (\frac{\partial z}{\partial t})\partial X^{i} - (\frac{\partial \bar{z}}{\partial t})\bar{\partial} X^{l} \\
&= (\frac{\partial z}{\partial t})\partial Y^{i} - (\frac{\partial \bar{z}}{\partial t})\bar{\partial} Y^{l} \quad . \\
&= \partial Y^{l}
\end{aligned}
\tag{17}
$$

We do the open string mode expansion

$$
\begin{aligned}
X_R^{\mu} &= \frac{1}{2} X_0^{\mu} + i2A^2 p^i \ln z + iA \sum_{n\neq 0} \frac{z^{-n}}{n} \alpha_n^{\mu} \\
X_L^{\mu} &= \frac{1}{2} X_0^{\mu} + i2A^2 p^i \ln \bar{z} + iA \sum_{n\neq 0} \frac{\bar{z}^{-n}}{n} \alpha_n^{\mu}
\end{aligned}
\tag{18}
$$

Here $p^i = n_i / R_j$ and $A = \sqrt{\alpha' / 2}$ = constant, the dual coordinate $Y^l = X_R^l - X_L^i$ satisfies

$$
\begin{aligned}
Y^l(\sigma = \pi) - Y^i(\sigma = 0) &= 2\pi\alpha' p^l \\
&= 2\pi\alpha' n_i / R_i \\
&= 2\pi n_i R_l'
\end{aligned}
\tag{19}
$$

$Y^l(\sigma = \pi)$ and $Y^i(\sigma = 0)$ are identical points on dual torus. This show a contour connecting them for any pair of boundaries

$$
\begin{aligned}
Y_{B_2}^l - Y_{B_1}^i &= \int_C d\sigma^a \partial_a Y^l \\
&= i \int_C d\sigma^a \varepsilon_a^b \partial_b Y^l \\
&= \alpha' p_C^l = 2\pi\alpha' n_l / R_l \\
&= 2\pi n_l R_l' \\
&\approx 0.
\end{aligned}
\tag{20}
$$

Here, $p_C^{\mu}$ is the spacetime momentum flowing across contour. We can find that, if $k$ spatial dimensions are compactified, and $R_i \to 0$ ($R_i$ is the strings' radius). We obtain the $(26-k)$ dimensions surface $Y^i$ that the open strings' end points are confined in ref[14].

Under the conformal geometic view,we decompose the surface $Y^i$ into the union set of the fundamental domain $D^{\mu,\nu,\cdots}$ and the cut graph $G^{\mu,\nu,\cdots}$. And the union set is a string $\gamma$ which the homotopy in the fundamental domain $D^{\mu,\nu,\cdots}$ and the boundary $\partial D^{\mu,\nu,\cdots}$. By the Seifert-van kampen theorem, we can shown by the following

$$\begin{aligned} \pi_1(D^{\mu,\nu,\cdots}) &= \langle e\rangle \\ \pi_1(G^{\mu,\nu,\cdots}) &= \langle \gamma_1, \ \dots \ ,\gamma_k\rangle \\ \pi_1(D^{\mu,\nu,\cdots}\cap G^{\mu,\nu,\cdots}) &= \langle\gamma\rangle. \end{aligned} \tag{21}$$

And

$$J{:}G^{\mu,\nu,\cdots}\cap D^{\mu,\nu,\cdots}\to G^{\mu,\nu,\cdots} \tag{22}$$

Therefore,

$$\pi_1(G^{\mu,\nu,\cdots}\cup D^{\mu,\nu,\cdots}) = \langle\gamma_1, \ \dots \ ,\gamma_k|J(\gamma)\rangle. \tag{23}$$

We calculate the fundamental group of the surface $Y^i$. We stipulate

$$\begin{aligned} I{:}W^{\mu,\nu,\cdots} &\to D^{\mu,\nu,\cdots} \\ J{:}W^{\mu,\nu,\cdots} &\to G^{\mu,\nu,\cdots} \end{aligned} \tag{24}$$

are the contains mapping. Choose a set of basis point $p^{\mu,\nu,\cdots}\in W^{\mu,\nu,\cdots}$ ($W^{\mu,\upsilon,\cdots}= D^{\mu,\upsilon,\cdots}\cap G^{\mu,\upsilon,\cdots}$) And then, each subset of the fundamental group are

$$\begin{aligned} \pi_1(G^{\mu,\nu,\cdots},p^{\mu,\nu,\cdots}) &= \langle G_1^{\mu,\nu,\cdots}, \ \dots \ ,G_k^{\mu,\nu,\cdots}|\alpha_1^{\mu,\nu,\cdots}, \ \dots \ ,\alpha_l^{\mu,\nu,\cdots}\rangle \\ \pi_1(D^{\mu,\nu,\cdots},p^{\mu,\nu,\cdots}) &= \langle D_1^{\mu,\nu,\cdots}, \ \dots \ ,D_m^{\mu,\nu,\cdots}|\beta_1^{\mu,\nu,\cdots}, \ \dots \ ,\beta_n^{\mu,\nu,\cdots}\rangle \\ \pi_1(W^{\mu,\nu,\cdots},p^{\mu,\nu,\cdots}) &= \langle W_1^{\mu,\nu,\cdots}, \ \dots \ ,W_p^{\mu,\nu,\cdots}|\gamma_1^{\mu,\nu,\cdots}, \ \dots \ ,\gamma_q^{\mu,\nu,\cdots}\rangle. \end{aligned} \tag{25}$$

So, the fundamental group of $Y^i$ is

$$\begin{aligned} \pi_1(Y^i,p^{,\mu,\nu,\cdots}) = \langle & G_1^{\mu,\nu,\cdots}, \ \dots \ ,G_k^{\mu,\nu,\cdots}, \ \dots \ ,D_1^{\mu,\nu,\cdots}, \ \dots \ ,D_m^{\mu,\nu,\cdots}| \\ & \alpha_1^{\mu,\nu,\cdots}, \ \dots \ ,\alpha_l^{\mu,\nu,\cdots},\beta_1^{\mu,\nu,\cdots}, \ \dots \ ,\beta_n^{\mu,\nu,\cdots}, \\ & I(W_1^{\mu,\nu,\cdots})J(W_1^{\mu,\nu,\cdots})^{-1}, \ \dots \ ,I(W_p^{\mu,\nu,\cdots})J(W_p^{\mu,\nu,\cdots})^{-1}\rangle. \end{aligned} \tag{26}$$

In terms of the **Eq**. (10), we have

$$\pi_1(Y^1\cup \ \dots \ \cup Y^i) = \langle a|a^i\rangle = \mathbb{Z}_i. \tag{27}$$

Whatever the number $i$ take, by the **Eqs**. (11) and (12), we find

$$\partial S = \partial Y^i = \partial X^i = 0 . \tag{28}$$

This equation is exactly equivalent to **Eq**. (17). This also establishes the relationship between conformal geometry and D-brane, closed and open strings. What we should do next is to talk about the relationship between them[21].

## 3.2 The relationship between D-branes closed & open strings

We consider the open strings and closed strings on D-branes. If the strings on the D-branes have the same two endpoints[13]. And they can convert each other, then we say they are homotopic with each other[6]. Therefore, we should pick a base point $p$ on the D-branes. We look at the closed strings which from the base point and back to the base point. And to classify them. When two closed strings are connected to each other, we can a multiplication of closed strings, and them form a longer closed strings. So the longer closed strings are the product of two closed strings

$$\gamma_0 \cdot \gamma_1 . \tag{29}$$

The closed strings that can contract into basis points are called unit element $e$ .

$$C(\gamma) = e \tag{30}$$

where $C$ is contraction operation, $e$ is the unit element.

Taking the closed string reversed, and reverse closed string is called the inverse of the closed string

$$[(\gamma_i)^{-1}]^{-1} = \gamma_i . \tag{31}$$

With this connected operation, we have a group which called the homotopy group of D-branes

$$\pi(S,p)$$

where $S$ is the surface, $p$ is the basis point. With the more concise representation that consider the generation of homotopy groups on D-branes:

$$\{a_1, b_1, a_2, b_2, \ \ldots \ , a_g, b_g\} \tag{32}$$

satisfy the following conditions

$$\begin{cases} a_i \cdot b_j = \delta_i^j \\ a_i \cdot a_j = 0 \\ b_i \cdot b_j = 0 \end{cases} \tag{33}$$

where $a \cdot b$ are represent number of the intersection with closed string $a$ and closed $b$, respectively, and $g$ is the genus on the D-branes. The generation of the D-branes homotopy group is called its canonical basis. Now we cut the D-brane along the canonical basis. The resulting surface is the basic domain of the D-branes, and the boundary of this fundamental domain can be reduced to a point

$$C\left(a_1 b_1 a_1^{-1} b_1^{-1} a_2 b_2 a_2^{-1} b_2^{-1} \ \ldots \ a_g b_g a_g^{-1} b_g^{-1}\right) = 0 . \tag{34}$$

Any closed string on the D-brane can be transformed by the homotopy, making it and the canonical basis only intersect with the base point.

Consider tow topological manifolds $M$ and $\tilde{M}$. And there is a continuous epimorphism $p\colon \tilde{M} \to M$. For any point $q \in M$, there must be an open set $U_q$ and the mapping $p^{-1}(U_q) = \bigcup_{\alpha} \tilde{U}_\alpha$ satisfied $\tilde{U}_\alpha \cap \tilde{U}_\beta \neq \varnothing$. And if the mapping $p|\tilde{U}_\alpha : \tilde{U}_\alpha \to U_q$ is topological homeomorphic, there is a overlay mapping $p : \tilde{M} \to M$. We call $M$ bottom space and $\tilde{M}$ overlay space.

Now we assume that D-brane exists but not unique. There are tow membranes $\tilde{M}_D$ and $M_D$. Consider a circuit $\gamma\colon [c, c]$ ($a$ is the real number) on $M$ and a path $\gamma\colon [c, d]$ ($c, d$ is the real number and $c \neq d$). So, exist a projection of path overlay membrane $\tilde{M}_D$ is equal to the circuit bottom membrane $M_D$

$$p \circ \tilde{\gamma} = \gamma . \tag{35}$$

Let us find a group of open cover in the bottom membrane $M_D$

$$\gamma \subset \bigcup_{i=0}^{k-1} U_i . \quad (36)$$

Then, find the original image of each open set in the overlay membrane $\tilde{M}_D$

$$\tilde{U}_{D_i} \in p^{-1}(U_{D_i}),\ \tilde{U}_{D_{i-1}} \cap \tilde{U}_{D_i} \neq \varnothing . \quad (37)$$

So there is a unique path

$$\tilde{\gamma} \subset \bigcup_{i=0}^{k} \tilde{U}_{D_i} . \quad (38)$$

We know that the open cover group has radius $U_{D_i}(r_i)$ and $\tilde{U}_{D_i}(r'_i)$. If we take the radius $r_i \to 0$ and $r'_i \to 0$, as the **Fig. 2** shown for that, we can get a projection between the closed string $\Gamma\colon [c,c]$ and open string $\tilde{\Gamma}\colon [c,d]$.

$$\lim_{r_i \to 0} \bigcup_{i=0}^{k-1} U_{D_i}(r_i) = \Gamma \quad (39)$$

$$\lim_{r'_i \to 0} \bigcup_{i=0}^{k-1} \tilde{U}_{D_i}(r'_i) = \tilde{\Gamma} . \quad (40)$$

By **Eq**. (35), we there must be a projection

$$p\colon \Gamma[c,c] \to \tilde{\Gamma}[c,d] . \quad (41)$$

It shown in **Fig. 3**

And this reveals a conclusion: there is some mappings *p* that makes it, when the D-brane at the bottom is lifting homotopy to the D-brane at the top, the closed strings at the bottom D-brane can be converted to the open strings at the top D-brane. In other words, when the D-branes are connected, the open strings and closed strings have intricate interplay in spacetime.

## 5 .Summarize and Outlook

In this paper, we can give somehow an intuitive explanation for open strings and closed strings operators by using the conformal geometry method in the topological string theory. With the comformal geometric method, we discussed the D-branes under non-dynamic spatiotemporal evolution. The results shown open strings and closed strings interact in spacetime, and D-branes undertake the work for connecting then. However, further work will be required to indentify D-branes dynamics. Using conformal geometry, we will explore the geometric equation of energy and mass on D-brane. And in combination with General Relativity to investigate more questions.

## Acknowledgments

We thank the Science and Technology Research Programof the Chongqing Municipal Education Commission (Grant No.KJ1712299 and KJ1712310) and Yangtze Normal University (GrantNo. 2016XJQN28 and 2016KYQD11) for financial support.

## Figures and Tables Captions

**Fig. 1**

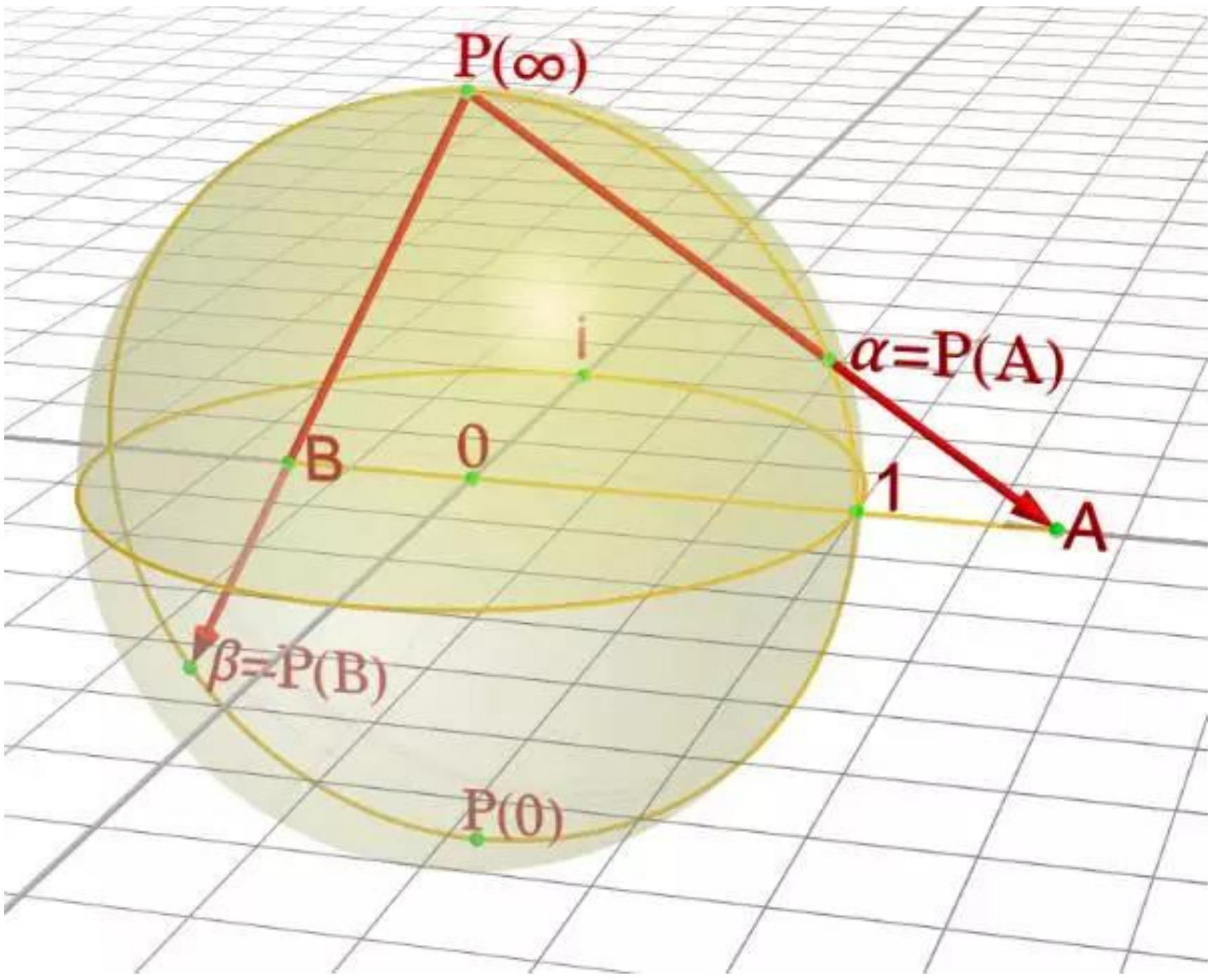

Fig. 1. Imagineing P(∞) is a light source in the north pole point. P(0) is the south pole point. A and B are points of intersection of light and equator. α and β are points of intersection of light and sphere.

**Fig. 2**

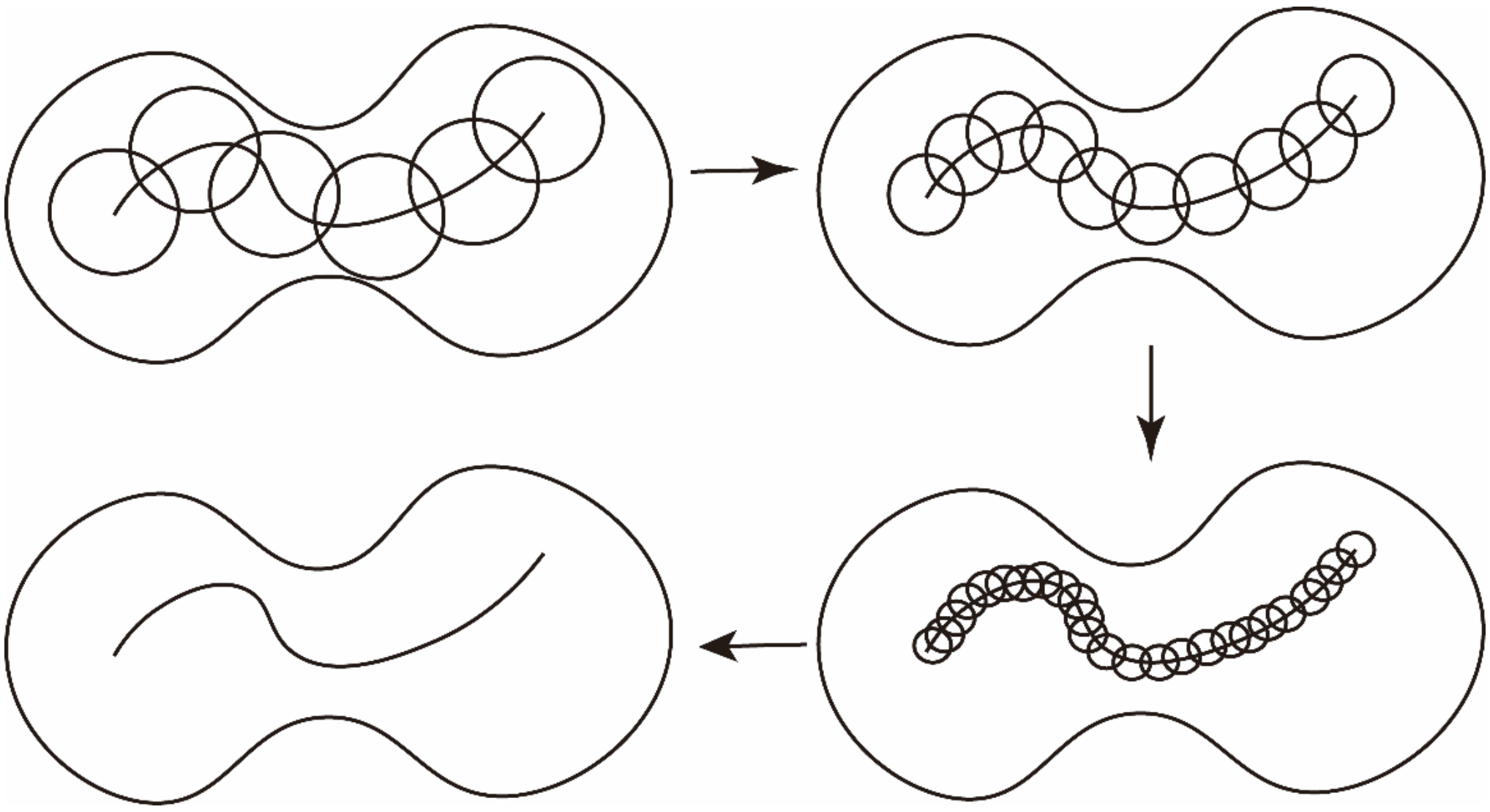

Fig. 2. When the $r_i$ (radius of the open cover) is contract to 0, the open string can be completely determined.

**Fig. 3**

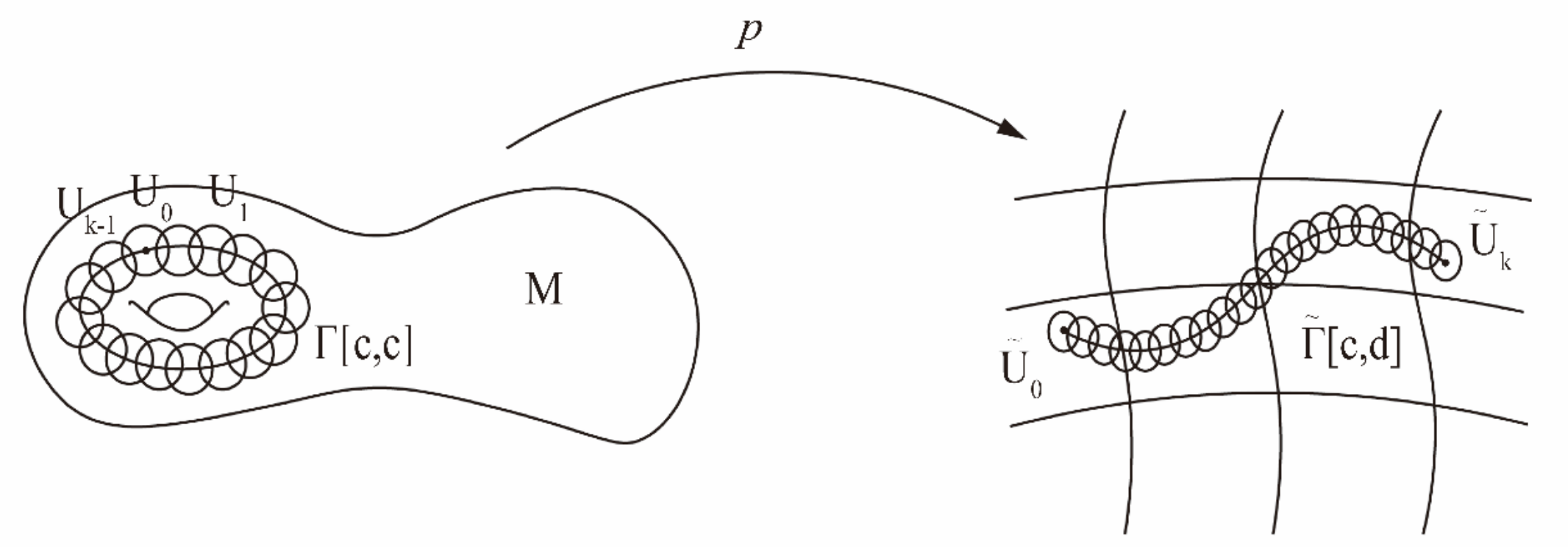

Fig. 3. $p{:}\Gamma[c,c]\to\tilde{\Gamma}[c,d]$ $p$ is the mapping which project the closed strings to the open strings.